# Metachronal motion of artificial magnetic cilia

Srinivas Hanasoge, Peter J. Hesketh, Alexander Alexeev

George W. Woodruff school of Mechanical Engineering,
Georgia Institute of Technology, Atlanta, GA, USA, 30332

Organisms use hair-like cilia that beat in a metachronal fashion to actively transport fluid and suspended particles. Metachronal motion emerges due to a phase difference between beating cycles of neighboring cilia and appears as traveling waves propagating along ciliary carpet. In this work, we demonstrate biomimetic artificial cilia capable of metachronal motion. The cilia are micromachined magnetic thin filaments attached at one end to a substrate and actuated by a uniform rotating magnetic field. We show that the difference in magnetic cilium length controls the phase of the beating motion. We use this property to induce metachronal waves within a ciliary array and explore the effect of operation parameters on the wave motion. The metachronal motion in our artificial system is shown to depend on the magnetic and elastic properties of the filaments, unlike natural cilia, where metachronal motion arises due to fluid coupling. Our approach enables an easy integration of metachronal magnetic cilia in lab-on-a-chip devices for enhanced fluid and particle manipulations.

Biological organisms use organelles such as hair-like cilia to perform vital functions involving fluid manipulation in their vicinity. Each cilium performs time-irreversible motion, traversing a spatially asymmetric path, and generates local fluid transport. It is often observed that carpets of natural cilia beat in a metachronal fashion, which appear as propagating waves. For example, lung cilia beat in a metachronal fashion to facilitate the net transport of fluid and microscopic particulates.[1–3] Similar metachronal motion is found on larger scales in the gait of millipedes, spiders and other invertebrates.[4]

In natural cilia, metachronal motion usually emerges as a self-organized phenomenon due to hydrodynamic coupling between a cilium and its neighbors.[3,5–7] As a result, tiny flexible cilia beat with a phase difference with respect to their neighbors, thereby generating travelling waves propagating along the ciliary carpet. The direction of metachronal waves in natural cilia can be symplectic (in the direction of the effective stroke), antiplectic (opposite to the direction of effective stoke), or laeoplectic (perpendicular to the direction of effective stoke).[8]

Studies point to specific advantages in the use of metachronal motion by natural cilia.[9,10] It has been reported that such motion can lead to a 3-fold increase in propulsion rate and a 10-fold increase in efficiency compared to synchronously beating cilia.[3] Metachronal beating is more energetically efficient and requires a decreased amount of adenosine triphosphate (ATP) consumed by cilia.[11] Furthermore, metachronal beating is advantageous in creating a unidirectional fluid pumping.[12] The attractive advantages offered by metachronal beating of natural cilia have motivated researchers to explore artificial analogs of ciliary carpets capable of performing metachronal motion that could be harnessed in microfluidic devices for efficient fluid transport and particle manipulation.[1–3,6,13,14]

To create metachronal motion artificially, it is essential to impose a phase difference in the beating cycles of multiple individual cilia. Such phase difference can be achieved by either applying different forcing to each cilium or by having cilia with different responses to a uniform forcing applied to the entire cilium array. The later approach that relies on a uniform actuation is more attractive from an experimental point of view. Indeed, large arrays of identical artificial cilia[15–19] have been previously actuated by a uniform magnetic force to explore microfluidic applications of their synchronous beating.[20–22] Only recently a uniform magnetic force was used by Tasumori et al.[23] to demonstrate metachronal motion in an array of artificial magnetic cilia that had a different magnetic orientation. The $2mm$ cilia were created using a sophisticated fabrication method that required carefully monitored processing steps with each cilium fabricated individually. The authors intend to miniaturize their cilia to a sub-millimeter size in the future work.

In this work, we explore a new approach to create metachronal motion in an array of magnetic cilia submerged in a viscous fluid and actuated by a uniform rotating magnetic field. High-aspect-ratio magnetic filaments that are periodically deformed by a rotating magnetic field have been previously suggested as an attractive approach to create artificial beating cilia.[24] We use cilia that are made up of paramagnetic metallic filaments with identical magnetic properties.[25,26] The motion of such cilia in a rotating magnetic field is defined by magnetic, elastic, and viscous hydrodynamic forces. The counter-clockwise rotating magnetic field bends the elastic cilium during the forward stroke, during which elastic energy is accumulated. After reaching maximum bending, cilia recover to the initial position by releasing the accumulated elastic energy. At relatively low actuation frequencies, the forward stroke is governed by a balance of the magnetic and elastic forces, whereas the recovery stroke is controlled by an interplay of the elastic and viscous forces.[19]

Metachronal wave motion in a ciliary array emerges when there is a phase difference in beating strokes of neighboring cilia, such that neighboring cilia transition from the forward to the recovery stroke in a sequential manner. In our system, the transition happens when the elastic force in deformed cilia exceeds the magnetic force due to the rotating magnet. More flexible cilia deflect more easily by the magnetic field[24] and, therefore, transition to the recovery at larger rotational angles of the magnet compared to stiffer cilia with the same magnetic properties. This leads to a phase difference between the beating of more flexible and less flexible magnetic cilia. In this work, we harness this property to fabricate an array of cilia that exhibit metachronal motion under uniform rotating magnetic field.

We consider elastic cilia that are made of a paramagnetic material. In this case, the ratio between magnetic and elastic forces acting on a cilium can be characterized by a magnetic number $Mn = BL\left(WP/\mu_0 EI\right)^{0.5}$. Here, $B = \left|\mathbf{B}\right|$ is the magnitude of the externally-applied magnetic flux density, $\mu_0 = 4\pi \times 10^{-7}$ is the permittivity of free space, $E$ is the Yung's modulus, $I = WP^3/12$ is the bending moment of inertia, and $L$, $W$, $P$ are the length, width, and thickness of the cilium, respectively.[25,27–29] Note that in the above definition of $Mn$, $W$ cancels out and, therefore, $Mn$ is independent of cilium width. Note also that a similarly defined magnetoelastic number has been previously successfully employed in numerical simulations to describe kinematics of magnetic cilia.[24,30,31]

To achieve metachronal motion, we can change the magnet angle at which cilia transition from forward to recovery stroke and, therefore, the phase of cilium beating, by altering the magnitude of $Mn$ for individual filaments in a ciliary array. Since $Mn$ linearly depends on the cilium length $L$, a linear change of cilium length should lead to a metachronal wave propagating along the ciliary array with a constant speed. To test this hypothesis, we fabricated a cilium array with cilium length increasing linearly with the cilium position along the array.

An array of nickel iron permalloy cilia is fabricated using surface micromachining with cilium length increasing from $60\,\mu m$ to $600\,\mu m$. The cilia are $10\,\mu m$ in width, and $60\,nm$ in thickness. The separation between neighboring cilia within a row is $50\,\mu m$, unless indicated otherwise. Multiple rows of cilia with separation $1\,mm$ are firmly attached to a glass slide and visualized using an inverted microscope (Nikon Eclipse Ti). Details of the fabrication process and imaging procedure can be found elsewhere.[25] The cilia are actuated by a permanent magnet with a $12\,mm$ diameter rotating with a constant frequency $f$. The large size of the magnet compared to the ciliary array ensures a uniform magnetic field experienced by cilia. A schematic of the experimental setup is shown in Fig. 1a.

Figures 1b, 1c, and 1d show snapshots of the ciliary array at different times during the magnet rotation. The snapshots illustrate the propagation of a metachronal wave along the array (also see Video 1 in ESI). We identify the front of the metachronal wave by the position of the cilia within the array that have just completed the recovery stroke, as indicated by the arrows in Figs. 1b, 1c, and 1d. We find that the metachronal wave propagates from the shorter cilia on the left to the longer cilia on the right with a nearly constant speed. This is indicated by the linearly increasing position of the wave front with time $T$ (Fig. 1e). Note that front position $X$ is normalized by total length of the array $L_A$. In our current implementation, the metachronal motion is perpendicular to the direction of the effective cilium stroke (i.e., laeoplectic).

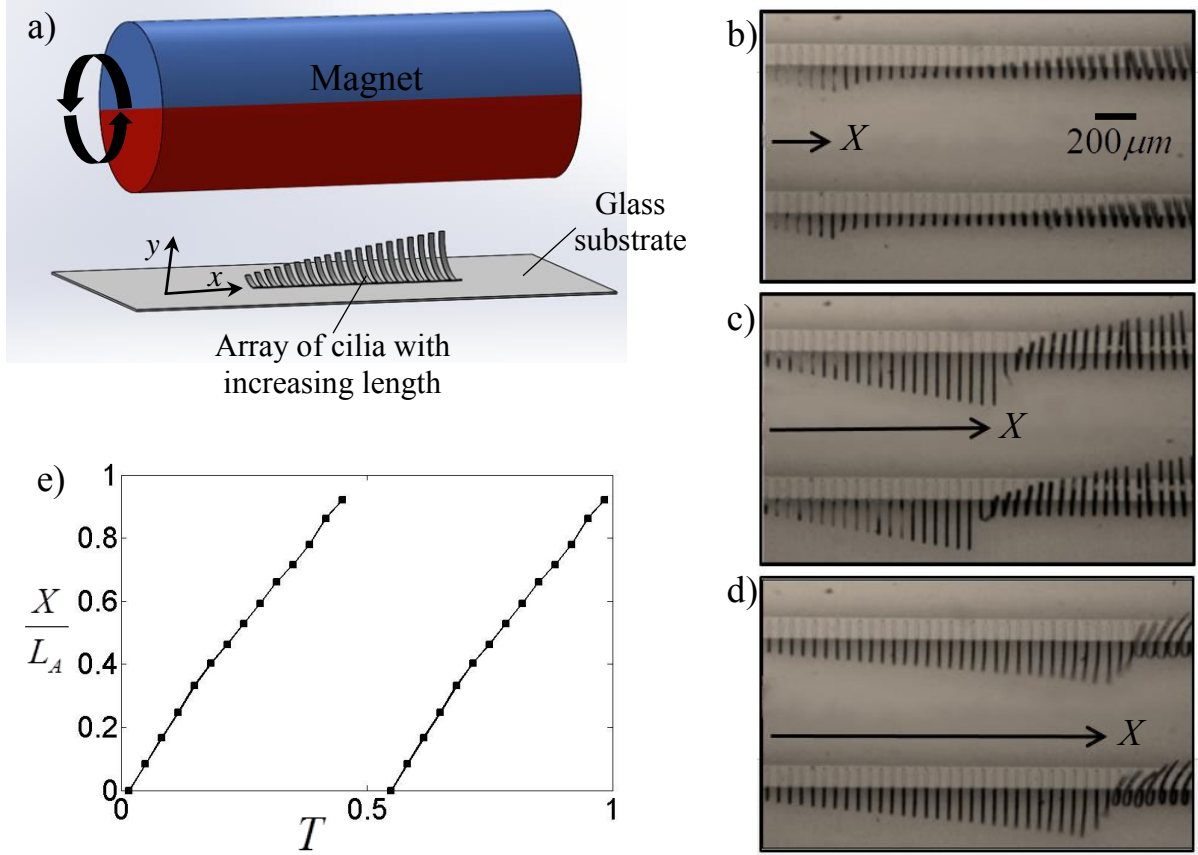

**Figure 1. a)** Schematic of the experimental setup with an array of cilia with linearly varying lengths actuated by a permanent magnet rotating. **b-d)** Snapshots of the ciliary array at time $T = 0.1$, $T = 0.25$, and $T = 0.45$, respectively. Two rows of cilia are shown. The magnet is rotated counter clockwise with a frequency of $0.5Hz$. Metachronal motion can be viewed from left to right from below the glass substrate, with the wave front indicated by the arrow. See ESI for videos of metachronal wave motion in ciliary arrays. **e)** Position of the metachronal wave front $\chi = X/L_A$, where $L_A$ is the array length, as a function of time $T$. Time $T$ is normalized by the period of magnet rotation.

To further understand how the metachronal beating is created in our ciliary array, we examine the kinematics of two beating cilia with different length. Figures 2a and 2b present a series of overlapped images of cilia with length $L = 220\mu m$ and $L = 480\mu m$, respectively. In both cases, cilium tips follow closed trajectories shown by a yellow line during the forward stroke and by a red line during the recovery stroke.

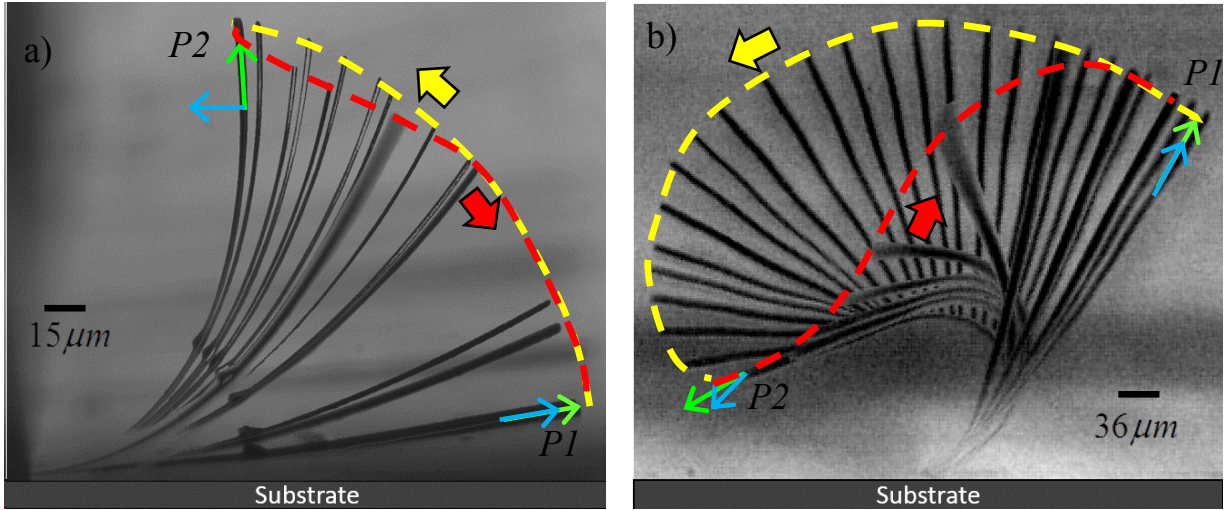

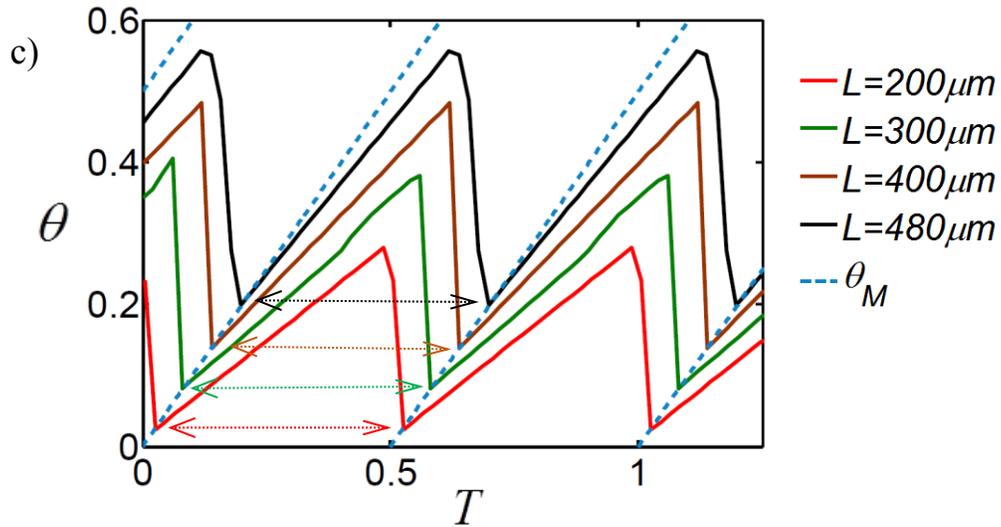

**Figure 2.** Motion of a single cilium in a beating cycle with length **a)** $L = 200\mu m$ and **b)** $L = 480\mu m$. The yellow and red arrows indicate the forward and recovery strokes, respectively. The green arrow is tangent to the cilium tip and the blue arrow indicates the direction of the magnetic field. **c)** The tip angle of beating cilia as a function of time $T$. The dotted lines show the rotational angle of the magnetic field. The angles are measured with respect to the substrate and are normalized by $2\pi$. Time is normalized by the period of the external magnetic field rotation equal to $2s$.

The forward stroke starts at position $P1$. For both the cilia, the tip angle, indicated by the green arrow, is closely aligned with the direction of the magnetic field, indicated by the blue arrow. The CCW rotating field induces a magnetic moment that bends the cilia in the counter-clockwise direction. The cilium tip angle increases and reaches a maximum at position $P2$. At this position of the maximum bending, the elastic force due to cilium deformation exceeds the magnetic force causing the beginning of the recovery stroke. Position $P2$ is determined by the value of $Mn$. Shorter cilia with larger stiffness, and therefore a lower $Mn$, bend to a lesser extent (Fig. 2a), whereas longer cilia with lower stiffness and higher $Mn$ are capable of bending to larger angles to follow the field (Fig. 2b). Thus, the maximum bending angle at position $P2$ is greater for the longer cilia that are characterized by higher values of $Mn$.

At position $P2$, the angle between the cilium tip and magnetic field is larger for the shorter cilium compared to the longer cilium, as indicated by the angle between the red and blue arrows in Fig. 2. Further rotation of the magnet beyond $P2$ reduces the magnetic moment and the cilia return to position $P1$ releasing the accumulated elastic energy. The shorter cilium returns to $P1$ at a smaller magnet angle than the longer cilium leading to a phase difference in the beating cycle that is proportional to the cilium length. Note that magnetization of the cilia flips its direction during the recovery stroke. As a result, cilia perform two beating cycles for each rotation of the magnetic field.[25,27,32]

The phase offset in the motion of different length cilia is illustrated in Fig. 2c. Here, we plot the normalized tip angle $\theta$ as a function of time $T$. Tip angle $\theta$ is measured with respect to the substrate in the counter-clockwise direction. Time $T$ is normalized by period of the magnetic field. The increasing value of $\theta$ indicates the forward stroke, whereas the decreasing $\theta$ represents the recovery stroke. The angle of the magnetic field $\theta_M$ is indicated by the tilted dotted lines.

The forward stroke starts when the direction of the magnetic field $\theta_M$ coincides with the tip angle $\theta$. This happens at different $T$ for cilia with different length, with shorter cilia initiating the forward stroke earlier than longer cilia. The tip angle increases nearly linearly with $T$ until the maximum is reached. As cilia deform, the difference between the cilia tip angle $\theta$ and the direction of magnetic fields $\theta_M$ increases, indicating the lag in cilium tip angle with respect to the magnet. This difference is larger for shorter cilia which are effectively stiffer. The maximum $\theta$ coincides the beginning of the recovery stroke, which starts earlier for shorter cilia. During the recovery stroke, $\theta$ decreases until it matches the direction of the magnetic field and then the cycle repeats. Thus, cilia of different length perform cyclic motion with period matching the half period of the magnet rotation, but with different phases with respect to the rotation of the magnetic field. The phase is proportional to the effective bending elasticity of the cilia and, thus, depends on the magnitude of the magnetic number $Mn$. Remember, $Mn = BL\left(WP/\mu_0 EI\right)^{0.5}$ represents the ratio between the magnetic force that is constant in our experiments and the elastic forces that is proportional to the cilium length. As a result, a metachronal wave emerges that propagates in the direction of the increasing cilium length (see Video 2 in ESI that shows two metachronal waves simultaneously propagating in opposing directions within a ciliary array with different cilium lengths).

In addition to the magnetic number $Mn$, cilium beating is characterized by the Reynolds number $Re = \rho L W f/\mu$ and the sperm number $Sp = L\left(\omega\xi/EI\right)^{0.25}$.[25,29,32] Here, $\rho$ is the fluid density, $\omega = 2\pi f$ is the cilium angular velocity, and $\xi = 4\pi\mu$ is the lateral drag coefficient of the cilium with $\mu$ being the fluid dynamic viscosity. The Reynolds number represents the importance of inertial hydrodynamic forces. In our experiments, the longest cilium length is $L = 600\,\mu m$ and the fastest actuation frequency is $f = 4Hz$, leading to the maximum $Re = 0.02$, indicating that fluid inertia has a negligible effect on cilium motion.

The sperm number $Sp$ characterizes the ratio of viscous to elastic forces acting on cilia. To further explore the effect of $Sp$ and $Mn$ on metachronal waves, we conducted experiments in which we varied both these dimensionless parameters in the ranges from 0 to 6.5 and from 0 to 4.9, respectively. Since, $Sp$ and $Mn$ are both proportional to $L$, their magnitudes change along our ciliary arrays. We, therefore, introduce dimensional parameters $Sp_L = Sp/L$ and $Mn_L = Mn/L$ to describes the operating condition for the entire array. In the experiments, we vary $Sp_L$ by changing the magnet frequency, whereas $Mn_L$ is changed by altering the distance from the array to the magnet. Note that $Sp_L$ and $Mn_L$ have the unit of $m^{-1}$.

To characterize the metachronal motion, we measure the time $T_c$ at which a cilium returns to $P1$ completing a beating cycle (see Fig. 2). In Fig. 3, we plot $T_c$ as a function of the normalized cilium position $\chi = X/L_A$, where $X$ is the cilium position within the array and $L_A = 3mm$ is the length of the array. The results are presented for selected values of $Sp_L$ and $Mn_L$. Figure 3 shows that $T_c$ increases nearly linearly with $\chi$ along the array indicating the propagation of a metachronal wave with a constant speed. Furthermore, the inverse of $T_c$ slope represents the wave speed.

We find in Fig. 3 that all the data for $T_c$ collapses into two curves corresponding to the two values of $Mn_L$ tested in our experiments. The results are independent of $Sp_L$ confirming that the metachronal wave motion is solely defined by $Mn_L$ that sets the speed of the wave propagation. A lower value of $Mn_L$ results in a faster wave speed. Indeed, weaker magnetic field can deform cilia to a smaller angle triggering an earlier transition to the recovery stroke. Therefore, $T_c$ decreases with decreasing $Mn_L$.

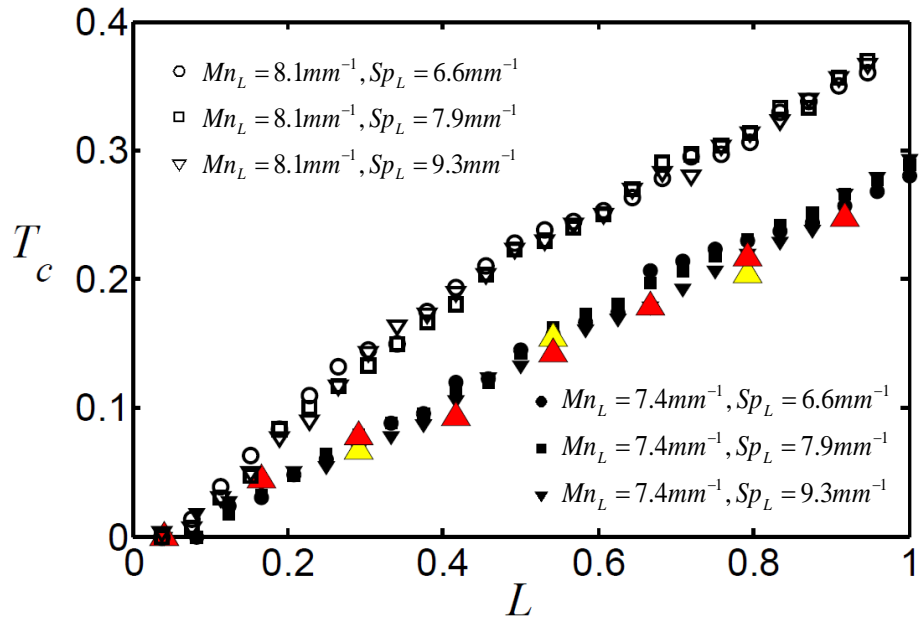

**Figure 3.** Cycle completion time $T_c$ as a function of cilium position $\chi$ for various experimental conditions. Note that data points collapse onto separate curves depending on $Mn_L$, indicating the weak dependence on $Sp_L$. The symbols ▲ and ▲ represent data obtained for $Mn_L = 7.4 mm^{-1}$, $Sp_L = 6.6 mm^{-1}$ from arrays with $150 \mu m$ and $300 \mu m$ spacing between cilia, respectively.

In naturally occurring cilia, metachronal motion emerges as a result of hydrodynamic coupling between neighboring cilia.[3,5–7] To probe the effect of hydrodynamic interactions between neighboring cilia in our synthetic system, we conducted experiments with ciliary arrays having different distances between cilia within a row. Specifically, we fabricated arrays of cilia with 3 and 6 times greater spacing between the neighboring filaments as compared to our original arrays. The results for $T_c$ obtained in these experiments are indicated in Fig. 3 by the colored symbols. We find that spacing between cilia does not affect metachronal wave propagation. Thus, in our artificial ciliary system the metachronal motion is solely controlled by the interplay between magnetic and elastic forces, and not due to hydrodynamics coupling.

In our experimental system, the metachronal wave propagates in the direction perpendicular to cilium beating plane, resulting in laeoplectic metachronal motion. Such laeoplectic metachronal motion has been shown to produce secondary flows with a net fluid flux perpendicular to the beating plane.[33] We note that by changing the spatial arrangement of cilia on the substrate such that cilia with different $Mn$ are placed in front or behind of each other, either symplectic or antiplectic motion can be achieved. It has been shown that for antiplectic metachrony, the net flow generated by cilia is greater, whereas symplectic metachronal beating leads to a flow that is slower in comparison to synchronously beating cilia.[33] Studying the fluid flow produced by magnetic cilia arranged in these different spatial configurations that lead to different cases of metachrony is an important direction of the future investigation that will also enable better understanding of the function of natural cilia.

In conclusion, we have demonstrated a new approach to create metachronal waves in an array of artificial cilia submerged in a viscous fluid and actuated by a uniform rotating magnetic field. We use arrays of metallic thin film cilia with varying length and make use of a phase difference in their motion to create periodical traveling waves propagating along the ciliary arrays in the direction of cilium length gradient. We show that the phase of a cilium beating is fully defined by the magnitude of the magnetic number that relates the strength of the magnetic force acting on cilia and cilium elasticity. For a given magnetic flux density, the magnetic number is proportional to cilium length, enabling the use of cilium geometry to induce metachronal motion. We show that this metachronal motion is insensitive to the magnitude of the viscous forces acting on beating cilia and hydrodynamic interactions between neighboring cilia. We note that the magnetic number depends on cilium thickness, but is independent of the cilium width. Thus, we expect that arrays of cilia with identical length and a gradient in cilium thickness will yield a similar metachronal behavior. More generally, metachronal waves can be expected in ciliary arrays with gradients of the magnetic number actuated by a uniform magnetic field. Finally, we note that our fabrication approach enables creation of large arrays of microscopic magnetic cilia that can be readily integrated into various microfluidic devices. This in turn will facilitate the development of attractive biomimetic platforms that use metachronal cilium motion for fluid and particle manipulation.

**Acknowledgments**

We thank the USDA NIFA (grant #11317911) and the National Science Foundation (CBET-1510884) for financial support and the staff of Georgia Tech IEN for assistance with clean room fabrication.